\documentclass[reprint,prl,showpacs,aps,amsmath,amssymb,nofootinbib,floatfix,floats,
superscriptaddress,twocolumn]{revtex4-1} 

\allowdisplaybreaks[2]
  
\usepackage[usenames, dvipsnames]{color}
\usepackage[colorlinks, pdfborder={0 0 0}, plainpages=false]{hyperref}
\usepackage{breakurl}
\usepackage{amsmath, amssymb, amsfonts}
\usepackage{xspace} 
\usepackage{dcolumn}
\usepackage{bm}
\usepackage{multirow} 
\usepackage{graphics,psfrag}
\usepackage{graphicx,psfrag}
\usepackage{natbib}
\def\laq{\raise 0.4ex\hbox{$<$}\kern -0.8em\lower 0.62ex\hbox{$\sim$}}
\def\gaq{\raise 0.4ex\hbox{$>$}\kern -0.7em\lower 0.62ex\hbox{$\sim$}}

\definecolor{CiteColor}{rgb}{0, 0.5, 0}
\hypersetup{citecolor=CiteColor} %
\definecolor{RefColor}{rgb}{0.55, 0, 0}
\hypersetup{linkcolor=RefColor} %

\usepackage{ulem}
\usepackage{amsmath}
\usepackage{times}
\usepackage{mathptmx}
\definecolor {darkgreen}{rgb}{0.2, 0.7, 0.2}

\newcommand{\Maryland}{\affiliation{Maryland Center for Fundamental
    Physics \& Joint Space-Science Institute,\\ Department of Physics,
    University of Maryland, College Park, MD 20742, USA}}
\newcommand{\Caltech}{\affiliation{Theoretical Astrophysics 350-17,
    California Institute of Technology, Pasadena, CA 91125, USA}}
\newcommand{\Cornell}{\affiliation{Center for Radiophysics and Space
    Research, Cornell University, Ithaca, New York 14853, USA}}
\newcommand{\CITA}{\affiliation{Canadian Institute for Theoretical
    Astrophysics, 60 St.~George Street, University of Toronto,
    Toronto, ON M5S 3H8, Canada}} %
\newcommand{\CIFAR}{\affiliation{Canadian Institute for Advanced Research, 180 Dundas St.~West, Toronto, ON M5G 1Z8, Canada}} %

\newcommand{\Fullerton}{\affiliation{Gravitational Wave Physics and Astronomy Center, California State University Fullerton, Fullerton, CA 92831, USA}}

\newcommand{\vS}{\mbox{$\bm{S}$}}
\newcommand{\vL}{\mbox{$\bm{L}$}}

\begin{document}

\title{Effective-one-body model for black-hole binaries with generic mass ratios and spins}

\author{Andrea Taracchini} \Maryland %
\author{Alessandra Buonanno} \Maryland %
\author{Yi Pan} \Maryland %
\author{Tanja Hinderer} \Maryland %
\author{Michael Boyle} \Cornell 
\author{Daniel A. Hemberger} \Cornell \Caltech
\author{Lawrence E. Kidder} \Cornell 
\author{Geoffrey Lovelace} \Fullerton \Caltech
\author{Abdul H.~Mrou\'{e}} \CITA 
\author{Harald P.~Pfeiffer} \CITA \CIFAR 
\author{Mark A.~Scheel} \Caltech 
\author{B\'{e}la Szil\'{a}gyi} \Caltech 
\author{Nicholas W.~Taylor} \Caltech
\author{Anil Zenginoglu} \Caltech

\begin{abstract}
Gravitational waves emitted by black-hole binary systems have the
highest signal-to-noise ratio in LIGO and Virgo detectors when
black-hole spins are aligned with the orbital angular momentum and
extremal.  For such systems, we extend the effective-one-body
inspiral-merger-ringdown waveforms to generic mass ratios and spins 
calibrating them to 38 numerical-relativity
nonprecessing waveforms produced by the SXS Collaboration. 
The numerical-relativity simulations span mass ratios from
1 to 8, spin magnitudes up to 98\% of extremality, and last for 40 to
60 gravitational-wave cycles.  When the total mass of the binary is
between $20M_{\odot}$ and $200M_{\odot}$, the effective-one-body
nonprecessing (dominant mode) waveforms have overlaps above $99\%$
(using the advanced-LIGO design noise spectral
density) with all of the 38 nonprecessing numerical waveforms, when
maximizing only on initial phase and time. This implies a negligible
loss in event rate due to modeling. Moreover, without further
calibration, we show that the precessing effective-one-body (dominant
mode) waveforms have overlaps above $97\%$ with two very long,
strongly precessing numerical-relativity waveforms, when maximizing
only on the initial phase and time.
\end{abstract}


\maketitle 

{\it Introduction.} In the next few years, second-generation ground-based interferometers, 
such as advanced LIGO~\cite{Shoemaker2009}, 
advanced Virgo~\cite{AdV} and KAGRA~\cite{Somiya:2011np}, 
will start to collect data with unprecedented sensitivity, making the long-sought detection of 
gravitational waves (GWs) a realistic prospect. Coalescing binaries of 
compact objects are among the most promising astrophysical sources 
in the accessible frequency band of such experiments. 
The search for GWs from these sources exploits
the matched-filtering technique, in which 
the noisy output of the interferometer is correlated with a bank of 
template waveforms describing all expected signals.  
An accurate knowledge of the gravitational radiation is thus crucial for
maximizing the chances of detection. However, matched-filtering not only
requires templates that are accurate, but their generation must also be
sufficiently cheap that they cover  the entire physical parameter space.
While in principle the most precise waveforms are obtained by solving
 Einstein's equations in numerical relativity (NR),
their considerable computational cost makes it necessary to resort
to analytical models that meet both criteria of accuracy and computational efficiency.

A unified analytical description of the entire compact binary coalescence, from the 
quasicircular inspiral, through the merger, and to the ringdown of the remnant, 
is achieved by the effective-one-body (EOB) model~\cite{Buonanno:1998gg, *Buonanno:2000ef}. 
In the EOB approach, one replaces the {\it real} problem of two compact objects of mass $m_{i}$, spin $\vS_{i}$ ($i=1,2$) and mass ratio $q$  orbiting each other with the {\it effective} 
problem of an extreme mass-ratio binary, where the more massive object
is a deformed-Kerr black hole (BH) and the small object is an effective 
spinning particle. The deformation parameter of the Kerr metric is 
the symmetric mass ratio $\nu \equiv q/(1+q)^{2}$. The EOB model incorporates  
results from post-Newtonian (PN) theory (in resummed form), BH 
perturbation theory, and more recently also from the gravitational self-force formalism. 
A mapping between the physical parameters of the
two problems is established by requiring that the effective dynamics is equivalent (when PN-expanded 
in powers of $1/c^2$) to the original, PN-expanded dynamics. Thus, solving exactly the 
effective problem of a spinning particle in the deformed-Kerr geometry 
amounts to introducing a particular {\it non-perturbative} method for resumming 
the PN-expanded equations of motion. 

The accuracy of the EOB waveforms has recently been improved by including in the EOB dynamics 
higher-order (yet unknown) PN terms and calibrating them to NR simulations, which have progressively 
grown in number, length and accuracy. State-of-the-art calibrations of these {\it adjustable} parameters
in the nonspinning sector (including also higher harmonics) can be found in
Refs.~\cite{Pan:2011gk,Taracchini:2012ig,Damour:2012ky}. An EOB model for spinning,
nonprecessing BH binaries was calibrated to 5 nonspinning and only 2
spinning, nonprecessing NR simulations in 
Ref.~\cite{Taracchini:2012ig}\footnote{The EOB models of Refs.~\cite{Buonanno:2007pf,Pan:2011gk,Taracchini:2012ig} 
have been implemented in the LIGO Algorithm Library under the names 
\texttt{EOBNRv1}, \texttt{EOBNRv2} and \texttt{SEOBNRv1}, 
respectively, and have been used in GW searches~\cite{Aasi:2012rja}.}; 
it can generate dominant (2,2) mode waveforms for any mass ratio, but only for BH spin magnitudes up to
0.6. Moreover, the EOB model in Ref.~\cite{Taracchini:2012ig} 
was compared and validated against a
large set of new NR simulations of nonprecessing BHs produced by several groups within the
numerical-relativity and analytical-relativity (NRAR) Collaboration~\cite{Hinder:2013oqa}. Recently,
Ref.~\cite{Pan:2013rra} provided a general procedure to generate EOB
waveforms for spinning, precessing BH binaries starting from a generic
spinning, nonprecessing EOB model; when using the EOB model in Ref.~\cite{Taracchini:2012ig} 
as the underlying nonprecessing model, the authors found remarkable agreement
with two precessing NR simulations. Finally, the conservative dynamics
of the EOB model has also been tested and validated through the study
of the periastron advance in BH binaries~\cite{LeTiec:2011bk,*Hinderer:2013uwa}. 

In this work, we calibrate the nonprecessing sector of a generic spinning EOB
model to the (2,2) mode of a catalog of highly-accurate NR simulations 
produced by the SXS Collaboration~\cite{Mroue:2013xna,*Mroue:2012kv,*Hemberger:2013hsa,*Hemberger:2012jz}. They include 
8 nonspinning and 30 spinning, nonprecessing BH binaries with
 spins up to 98\% of extremality,
they cover mass ratios up to 8, and have orbital eccentricities in the range of 
a few percent down to $10^{-5}$.
The simulations follow more orbits on average (up to 35.5),
 allowing a more reliable calibration of analytical waveforms. 

{\it Effective-one-body model.} In what follows we set $G=c=1$. Let $\hat{\vL}$ be
the direction perpendicular to the binary's instantaneous orbital plane, and let us define the
dimensionless projections of the spins along $\hat{\vL}$ as $\chi_{i}\equiv
(\vS_{i}\cdot \hat{\vL})/m_{i}^{2}$. We assume $m_{1}\geq m_{2}$,
hence $q\equiv m_{1}/m_{2}\geq 1$. In the spinning EOB formalism of 
Ref.~\cite{Barausse:2011ys}, the effective Hamiltonian $H_{\rm eff}$ is that 
of a particle of mass $\mu\equiv m_{1}m_{2}/(m_{1}+m_{2})$
and effective spin $\vS^{*}\equiv \vS^{*}(\vS_{1},\vS_{2})$ moving in a
deformed-Kerr geometry of mass $M\equiv m_{1}+m_{2}$ and spin $\vS_{\rm
  Kerr}\equiv \vS_{1}+\vS_{2}$; the conservative orbital dynamics is
then derived via Hamilton's equations using the real EOB-resummed Hamiltonian
\begin{equation}
\label{Hreal}
  H_{\text{real}}=M\sqrt{1+2\nu\left(\frac{H_{\text{eff}}}{\mu}-1\right)}-M\,.
\end{equation} 
We use here the same EOB Hamiltonian as in
Ref.~\cite{Taracchini:2012ig}, but augment the deformed-Kerr metric
  potential $\Delta_u$ with 4PN nonspinning terms to obtain~\cite{Barausse:2011ys}
\begin{eqnarray}
  \label{delta_u}
  \Delta_u(u) &=& \bar{\Delta}_u(u)\, \Big[1 + \nu\,\Delta_0 + 
    \nu \,\log \Big(1 + \sum_{i=1}^{5}\Delta_{i}u^{i}\Big)\Big]\,,
\end{eqnarray}
where $u\equiv 1/r$ and $r$ is the EOB radial coordinate in units of $M$. Here,
\begin{subequations}
  \begin{align}
    \bar{\Delta}_u(u)=&\,\chi_{\text{Kerr}}^2\,\left(u -
      \frac{1}{r^{\text{EOB}}_{+}}\right)\,
    \left(u - \frac{1}{r^{\text{EOB}}_{-}}\right)\,,\\
    \label{eq:hor}
    r^{\text{EOB}}_{\pm} =&\, \left[1\pm
      \left(1-\chi^2_\text{Kerr}\right)^{1/2}\right]\,(1-K\,\nu)\,,
  \end{align}
\end{subequations}
with $\chi_{\rm Kerr}\equiv (\vS_\text{Kerr}\cdot \hat{\vL})/M^{2}$;
the coefficients $\Delta_{0}, \ldots, \Delta_{5}$ are determined by
requiring that $\Delta_{u}$ agrees with the Taylor-expanded EOB
potential $A(r)$~\cite{Barausse:2011dq,Bini:2013zaa} up to 4PN order.  
By construction, $r^{\text{EOB}}_{\pm}$ reproduce the Kerr horizons when
$\nu=0$. Similarly to what was done in Ref.~\cite{Taracchini:2012ig},
we exploit $K$ as an adjustable parameter, i.e., a parameter that 
we calibrate to NR waveforms. For the identification between the effective particle's spin $\vS^{*}$
and the spins $\vS_{i}$  we use the 3.5PN-accurate spin 
mapping of Ref.~\cite{Barausse:2011ys}, 
with all the arbitrary gauge parameters set to zero and with the addition of a 4.5PN spin-orbit 
term of the form $(d_{\rm SO}\nu \vS_{\rm Kerr})/r^{3}$, where $d_{\rm SO}$ is an
adjustable parameter. The EOB description of conservative spin effects is completed by adding 
a 3PN spin-spin term of the form $d_{\rm SS}\nu
(\vS_{1}^{2}+\vS_{2}^{2})/r^{4}$ to $H_{\rm eff}/\mu$, where $d_{\rm SS}$ is another adjustable
parameter. 

The adjustable parameters are chosen to be polynomials in $\nu$ whose coefficients are determined 
by minimizing the phase and amplitude difference between EOB and NR waveforms 
via the numerical simplex method for each mass ratio. First, we calibrate the
nonspinning sector and find $K = 1.712 - 1.804\nu - 39.77\nu^{2} + 103.2\nu^{3}$, where the $\nu$-independent 
term is consistent with the frequency shift of the innermost stable circular orbit (ISCO) due
to conservative self-force effects in the small-mass-ratio 
limit~\cite{Barack:2010tm}. Next, we calibrate the spin parameters and obtain $d_{\rm SO}=-74.71 - 156.0\nu + 627.5 \nu^2$ and $d_{\rm SS} =8.127 - 154.2\nu + 830.8\nu^2$.

Dissipative effects are modeled by supplementing Hamilton's equations
with a radiation-reaction force which is a sum over (time derivatives
of) the $-2$-spin-weighted spherical modes at infinity. In our
model, these modes are written as a factorized resummation of the PN
waveforms~\cite{Damour:2008gu,Pan:2010hz}
\begin{equation}\label{hlm}
  h^\mathrm{F}_{\ell m}=h_{\ell m}^{(N,\epsilon)}\,\hat{S}_\text{
    eff}^{(\epsilon)}\, T_{\ell m}\, e^{i\delta_{\ell 
      m}}\left(\rho_{\ell m}\right)^\ell\,
\end{equation}
(see Ref.~\cite{Taracchini:2012ig} for the definition of the
individual factors). In particular, here we also include comparable-mass
spin-orbit and spin-spin effects up to 2PN order, using the most recent 
PN-waveform calculations in Ref.~\cite{Buonanno:2012rv}. We use the $\rho_{\ell m}$-factorization
in Eq.~(\ref{hlm}) for all modes except those with $\ell \leq 4$ and
odd $m$, which instead follow the prescription of
Ref.~\cite{Taracchini:2012ig} (see the discussion above Eq.~(A8a)
therein). In addition, we also include all the known spin effects from the test-particle limit 
given in Ref.~\cite{Pan:2010hz}, by replacing the Kerr spin parameter $a/M$ 
with $\chi_{\rm Kerr}$; this helps the modeling of unequal-mass, spinning systems. As such, the mode amplitudes contain no adjustable parameters. In fact, 
the improved knowledge of the nonspinning sector (i.e., the addition of 4PN 
terms in $\Delta_{u}$) allowed us to remove the nonspinning adjustable parameter 
$\rho_{22}^{(4)}$ which had been introduced in Ref.~\cite{Taracchini:2012ig}, 
thus simplifying the nonspinning model. The resulting residuals on the amplitude of the (2,2)
mode are within a few percent at merger for $\chi_{1,2}\sim1$ even
without adding non-quasicircular corrections. However, we need to introduce an 
adjustable parameter in the spin terms of the phase $\delta_{22}$ to
enhance the EOB GW frequency close to merger with respect to its
leading-order value (twice the orbital frequency $\Omega$),
which tends to underestimate the NR value for $\partial_{t}\phi_{22}$ 
when spins are close to 1. For $\chi_{1,2}=0.98$, we find 
that the ISCO is crossed only $10M$ before the light ring crossing, 
thus greatly reducing the region in which the non-quasicircular corrections (see below)
can be effective. Explicitly, if $\chi \geq0$, we add the 3PN term 
$540\, \nu \chi (M\Omega)^{2}$ to $\delta_{22}$, where $\chi\equiv \chi_{\rm S}+\chi_{\rm A}\sqrt{1-4\nu}/(1-2\nu)$, with $\chi_{\rm S,A}\equiv (\chi_{1}\pm\chi_{2})/2$. 

Non-quasicircular (NQC) effects that become important near the merger are included in $h^\mathrm{F}_{22}$
through a factor $N_{22}$ (see Eq.~(18) of
Ref.~\cite{Taracchini:2012ig}). 
The NQC coefficients are fixed by
requiring that the EOB (2,2) mode agrees with the NR input values for 
$|h_{22}|$, $\partial_{t}|h_{22}|$, $\partial^{2}_{t}|h_{22}|$,
$\partial_{t}\phi_{22}$ and $\partial^{2}_{t}\phi_{22}$, evaluated at
the peak of $|h_{22}|$. Using the 38 NR nonprecessing waveforms in the
SXS catalog and Teukolsky waveforms computed in the small-mass-ratio
limit~\cite{Taracchini2013}, we updated the fitting formulas for
the NR input values given in Table~IV of
Ref.~\cite{Taracchini:2012ig}. We use these to iteratively compute the NQC coefficients 
as described in Sect.~IIB of Ref.~\cite{Taracchini:2012ig}. While previous nonspinning 
EOB models~\cite{Buonanno:2007pf} were calibrated without enforcing any time delay 
between the peak in the (2,2) amplitude and in the orbital frequency, here, as in 
Refs.~\cite{Pan:2011gk,Taracchini:2012ig}, we require a lag $\Delta
t^{22}_{\rm peak}$ which varies with the physical parameters of the
binary. The idea of introducing $\Delta t_{\rm peak}^{22}$ into the
model was inspired by studies in the small-mass-ratio
limit, where such time delay was first seen
with EOB trajectories sourcing Teukolsky
waveforms~\cite{Barausse:2011kb} and accurately
quantified in Ref.~\cite{Taracchini2013}. 
Finally, the inspiral-plunge waveform
is simply defined as $h_{22}^{\rm insp-plunge}\equiv N_{22}h_{22}^{\rm
  F}$, and $h_{\ell m}^{\rm insp-plunge}\equiv h_{\ell m}^{\rm F}$ when
$(\ell,m)\neq(2,2)$.

\begin{figure}[!ht]
\begin{center}
\includegraphics[width=0.48\textwidth]{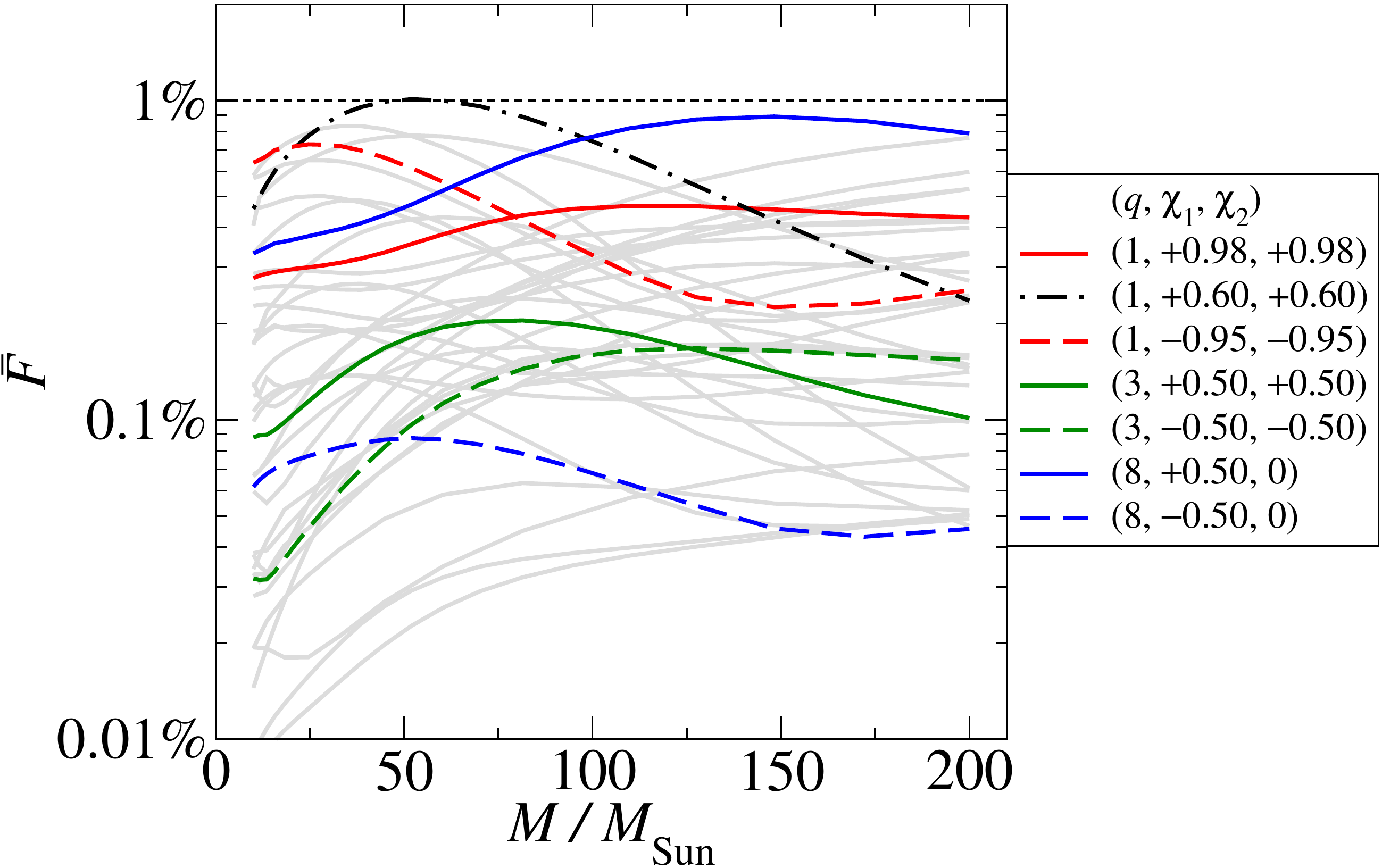}
\caption{\label{fig:Unfaith} Unfaithfulness of (2,2) EOB waveforms for \emph{all} the 38 nonprecessing BH binaries in the SXS catalog. Only a few selected cases are labeled in the legend.}  

\end{center}
\end{figure}

As usual, the EOB merger-ringdown (RD) waveform is built as a linear combination of 
quasi normal-modes (QNMs) of the remnant BH~\cite{Buonanno:2000ef}
\begin{equation}\label{ringdown}
  h_{\ell m}^{\rm merger-RD}(t)=\sum_{n=0}^{N-1}
  A_{\ell m n}\,e^{-i\sigma_{\ell m n}(t-t_{\text{match}}^{\ell m})}\,,
\end{equation}
where $N$ is the number of overtones, $t_{\rm match}^{\ell m}$ is the
time when $|h_{\ell m}^{\rm insp-plunge}|$ peaks, $A_{\ell m n}$ is
the complex amplitude of the $n$-th overtone of the $(\ell, m)$ mode,
and $\sigma_{\ell m n}=\omega_{\ell m n}-i/\tau_{\ell m n}$ is its 
complex frequency, having positive (real)
frequency $\omega_{\ell m n}$ and decay time $\tau_{\ell m n}$. The
frequencies $\sigma_{\ell m n}$ depend on the mass $M_f$ and
spin $a_f$ of the final Kerr BH, and are tabulated in
Ref.~\cite{Berti:2005ys}. To predict $M_{f}$ we use the
phenomenological formula proposed by Ref.~\cite{Barausse:2012qz}, but
we replace its equal-mass limit [Eq.~(11) therein] with the highly
accurate fit given in Eq.~(9) of Ref.~\cite{Hemberger:2013hsa}.  
To compute $a_{f}$, we start from the formula of 
Ref.~\cite{Barausse:2009uz} (which also predicts the direction of the
final spin for precessing binaries), and use the simulations in the SXS calatog to refit its nonprecessing limit; the main
change we introduce are 4 new fitting
coefficients designed to improve the equal-mass, high-spin corner of
the parameter space, where the prediction of Ref.~\cite{Barausse:2009uz} 
has residuals exceeding $5\%$. We improve the stability of the ringdown modeling across the entire parameter space by (i) assuming a monotonic behavior of $a_{f}$ with decreasing $\nu$ for extremal spins, and (ii) replacing some of the higher physical overtones with pseudo-QNMs that depend on the merger frequency, on $\sigma_{220}$ and on $\nu$, and moderate the rise of the ringdown GW frequency~\cite{Pan:2011gk,Taracchini:2012ig}.

Finally, the complete inspiral-merger-ringdown waveform is built as the
smooth matching of $h_{\ell m}^{\rm insp-plunge}$ to $h_{\ell m}^{\rm
  merger-RD}$ at $t_{\rm match}^{\ell m}$, over an interval $\Delta
t_{\rm match}^{\ell m}$, following the hybrid matching procedure of
Ref.~\cite{Pan:2011gk} to fix the coefficients $A_{\ell m n}$ in
Eq.~\eqref{ringdown}. 
\begin{figure}[!ht]
  \begin{center}
    \includegraphics*[width=0.45\textwidth]{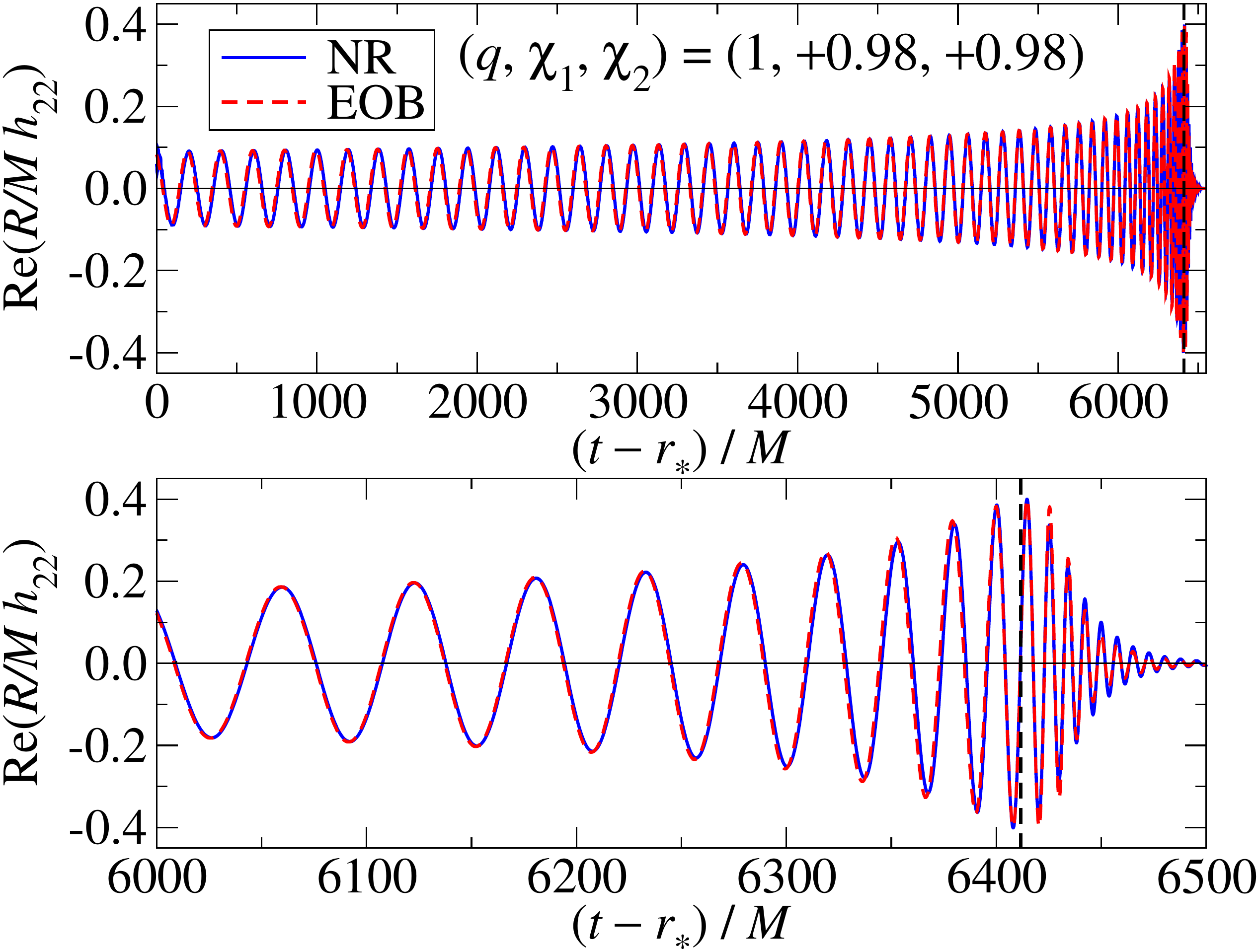}
    \caption{\label{fig:IMR098} NR and EOB (2,2) waveforms of the BH binary with $q=1$ and $\chi_{1}=\chi_{2}=0.98$. The two waveforms are aligned at their amplitude peak (marked by a vertical dashed line). 
$R$ is the distance to the source.}
  \end{center}
\end{figure}
\begin{figure}[!ht]
  \begin{center}
   \includegraphics*[width=0.43\textwidth]{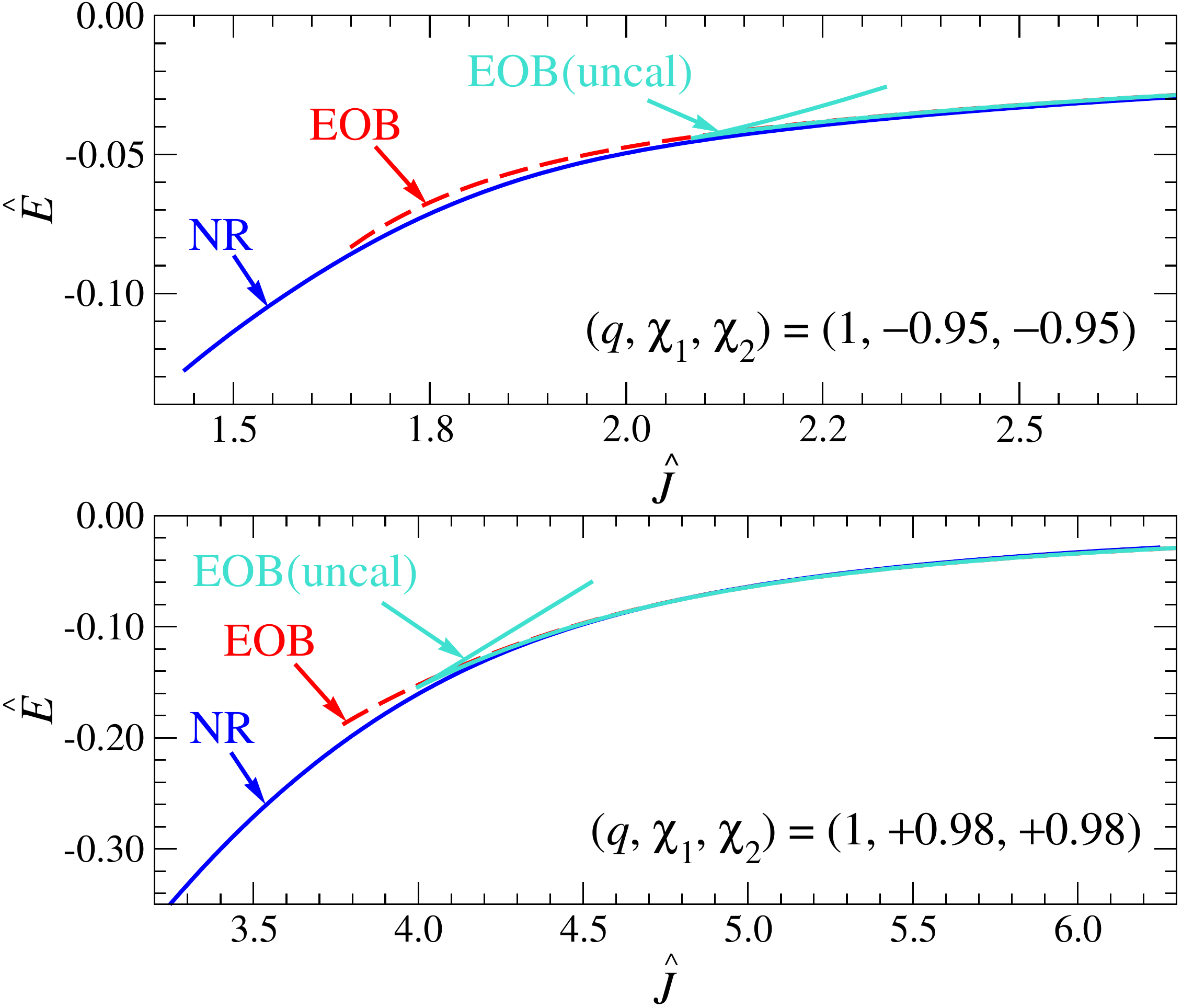}
    \caption{\label{fig:EJPaper} The specific binding energy $\hat{E}=E/\mu$ as a function of the dimensionless total angular momentum $\hat{J}=J/(\mu M)$ of the BH binaries with $q=1$ and $\chi_{1}=\chi_{2}=-0.95,\,0.98$ computed in NR, conservative uncalibrated EOB model and the calibrated EOB model of this paper.}
  \end{center}
\end{figure}

{\it Results and discussion.} The SXS catalog includes 8 nonspinning
BH binaries with $q=1$, 1.5, 2, 3, 4, 5, 6, 8, and 30 spinning,
nonprecessing BH binaries with: $q=1$ and $\chi_{1}=\chi_{2}=0.98$,
$0.97$, $\pm 0.95$, $\pm0.9$, $0.85$, $\pm 0.8$, $\pm0.6$, $\pm 0.44$,
$\pm 0.2$; $q=1,\, 1.5,\, 3,\, 5,\, 8$, $\chi_{1}=\pm0.5$ and
$\chi_{2}=0$; $q=1.5$ and $\chi_{1}=-\chi_{2}=\pm0.5$; $q=2$,
$\chi_{1}=0.6$ and $\chi_{2}=0$; $q=3$ and $\chi_{1}=\chi_{2}=\pm
0.5$. We find that to accurately match all 38 nonprecessing waveforms, it
  is sufficient to calibrate the EOB model to a much smaller subset of
  them. However, since our goal is an accurate model 
  for the entire parameter space, most of which is not covered by the 
  NR waveforms, we prefer to exploit all available non-degenerate NR
  information in the calibration. In Fig.~\ref{fig:Unfaith} we
compare the EOB waveforms to \emph{all} the 38 nonprecessing NR
waveforms by computing their unfaithfulness 
\begin{equation}
\bar{F}\equiv 1 - \max_{t_{0},\phi_{0}} \frac{\langle h_{22}^{\rm EOB},h_{22}^{\rm NR}\rangle}{||h_{22}^{\rm EOB}||\,||h_{22}^{\rm NR}||}\,,
\end{equation}
where $t_{0}$ and $\phi_{0}$ are the initial time and phase, $||h||\equiv\sqrt{\langle h,h\rangle}$, and the inner product between two waveforms is defined as $\langle h_1,h_2\rangle\equiv 4 {\rm Re} \int_{f_{\rm min}}^\infty \tilde{h}_1(f)\tilde{h}^*_2(f)/S_n(f)\,df$, where $S_n(f)$ is the
 zero-detuned, high-power noise spectral density of
advanced LIGO~\cite{Shoemaker2009} and $f_{\rm min}$ is the
starting frequency of the NR waveform (after junk radiation has
settled). We do not hybridize the NR waveforms at low frequency
($f<f_{\rm min}$) with any analytic approximant but instead taper
the EOB waveforms. When $M \,\laq\, 100 M_\odot$ the NR waveforms 
do not cover the entire frequency bandwidth of the detector, but 
we expect that the unfaithfulness $\bar{F}$ would not change much 
when longer NR waveforms will be employed because the EOB 
calibration has been shown to be quite stable with respect to 
the number of GW cycles used for the calibration~\cite{Pan2013b}. 
The unfaithfulness is always below $1\%$ for total masses from 
$20M_{\odot}$ to $200M_{\odot}$, implying a negligible loss in event rate
 due to the modeling error alone. 

\begin{figure}[!ht]
  \begin{center}
    \includegraphics*[width=0.45\textwidth]{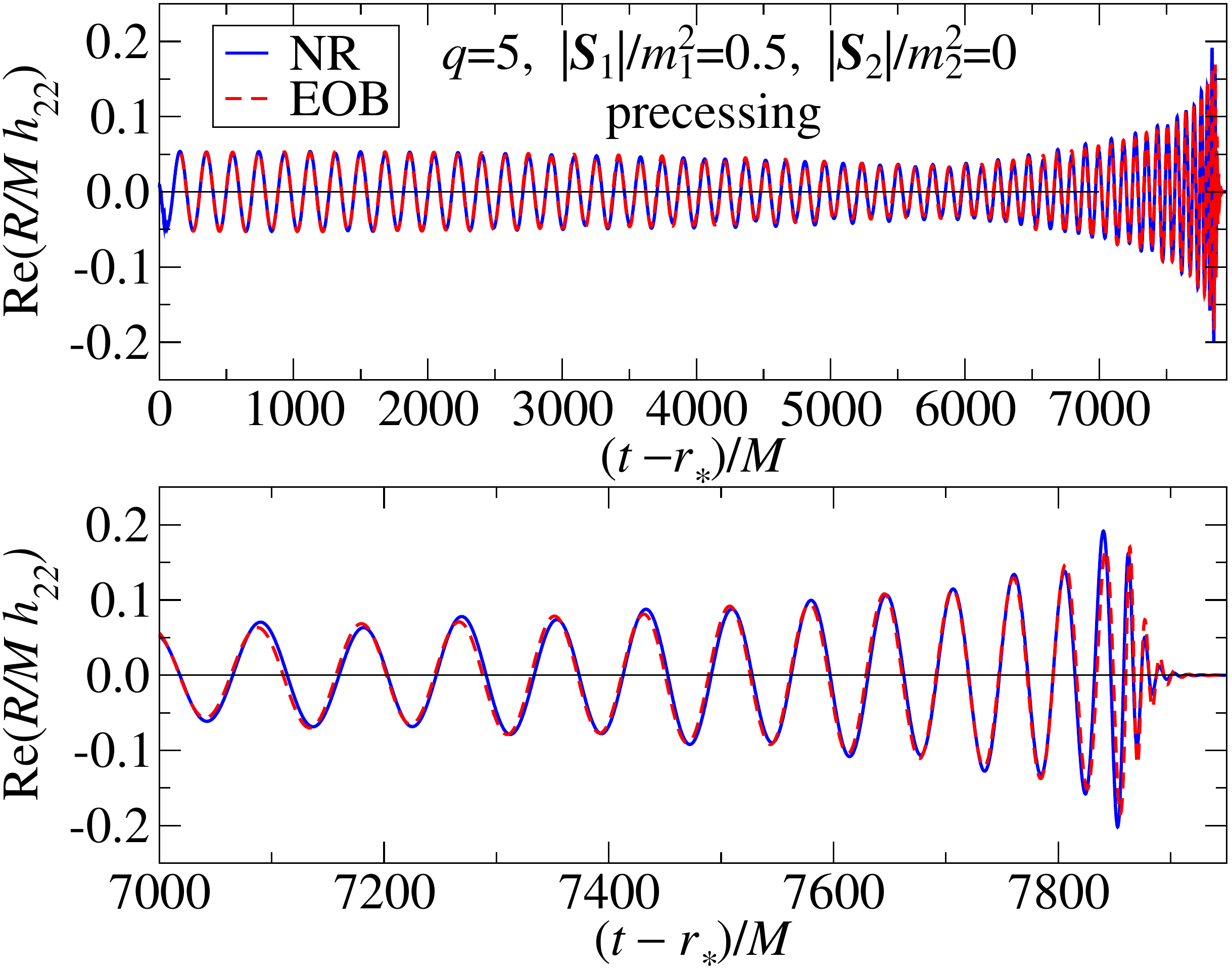}
    \caption{\label{fig:precessing} NR and EOB (2,2) precessing waveforms of 
the BH binary with $q=5$ and initial spins $|\vS_1|/m_{1}^{2}=0.5$ in the orbital plane and
$|\vS_2|/m_{2}^{2}=0$. The two waveforms are aligned at low frequency. $R$ is the distance to the source.}
  \end{center}
\end{figure}

To estimate the NR error for each binary configuration, we choose
the NR simulation with the largest number of
cycles, with the highest resolution, and extrapolated to infinity with
extrapolation order $N=3$ as the fiducial waveform. We then compute the model's unfaithfulness against NR waveforms: i) with a different extrapolation
order but the same resolution; ii) with a different resolution but
the same extrapolation order; and obtain a conservative error bound on
$\bar{F}$ from the difference between the fiducial and the
most discrepant waveform. For the binary with $q=1$ and
$\chi_{1}=\chi_{2}=0.98$, which we take as a representative case for the rest of the
catalog, the errors on $\bar{F}$ are within 0.005\%.

Figure~\ref{fig:IMR098} shows the agreement between EOB and NR
waveforms for the nearly extremal BH binary with $q=1$ and
$\chi_{1}=\chi_{2}=0.98$, when aligning them at their amplitude peak;
the phase difference is always within 0.6 rads. The coordinate
invariant relation $\hat{E}(\hat{J})$ between the specific energy
$\hat{E}$ and the total angular mometum $\hat{J}$ is a useful tool for
evaluating analytical descriptions of the binary
dynamics~\cite{Damour:2011fu,*LeTiec:2011dp}.  In
Fig.~\ref{fig:EJPaper}, for the cases with $q=1$ and
$\chi_{1}=\chi_{2}=-0.95,\,0.98$, we compare $\hat{E}(\hat{J})$ from
NR (using Cauchy-characteristics-extracted waveforms),  
the conservative uncalibrated EOB model, and the EOB model
calibrated in this paper. The numerical errors of $\hat{E}(\hat{J})$
  increase from $10^{-5}$ at low frequency to $10^{-4}$ at high
  frequency. We find that when the spins are close to extremal, there is a difference
  of $10^{-3}$ between NR and analytical (EOB or even PN) $\hat{E}(\hat{J})$ at low
  frequency that is not explained by numerical errors. By contrast the difference is
  within numerical errors when the spin magnitudes are less than $\sim
  0.6$. We plan to further investigate those results in the future. The cusps
in the conservative EOB curves indicate the presence of an ISCO; this
point lies $60M$ ($10M$) in time before merger for spin $-0.95$
(0.98). The calibrated EOB curves instead extend up to the light ring,
which is very close to the merger. The good agreement between EOB and
NR results validate the calibration procedure in yielding an accurate
description of the binary evolution up to merger.  The improved model
for the nonprecessing limit developed here (as compared to
Ref.~\cite{Taracchini:2012ig}) is also the foundation for precessing
binaries, via the procedure of transforming from the precessing frame
to an inertial frame described in Ref.~\cite{Pan:2013rra}. Without
further calibration, we tested our model against the 2 long precessing
waveforms that were used in Ref.~\cite{Pan:2013rra}, one with $q=3$
and initial spins (both of magnitude 0.5) respectively in the orbital
plane and antialigned with $\hat{\vL}$, and the other with $q=5$ and
initial spins (of magnitude 0.5 and 0, respectively) in the orbital
plane, and found that $\bar{F} < 3\%$ for both cases. We show EOB and
NR precessing waveforms of the $q=5$ case in
Fig.~\ref{fig:precessing}.

{\it Conclusions.}  Using 38 NR (2,2) mode waveforms for spinning,
nonprecessing BH binaries produced by the SXS Collaboration, we have
calibrated (with 27 of the 38 numerical waveforms) the nonprecessing sector of the EOB model of 
Refs.~\cite{Barausse:2011ys,Pan:2013rra}, which is valid for any 
mass ratio and spins. Throughout the entire parameter space covered by the NR simulations, 
the  EOB model of this paper achieves an unfaithfulness within $ 1\%$, implying a negligible loss 
in event rate due to the modeling error alone. By extending 
the EOB model to nearly extremal spins, we have increased the distance 
reach of advanced detectors. Furthermore, the EOB model can be used to generate 
precessing waveforms using the prescriptions in Ref.~\cite{Pan:2013rra}. 
The EOB model developed here will be implemented in the LIGO Algorithm Library, so that it can be 
employed by advanced LIGO and Virgo to detect gravitational-waves from spinning binary BHs 
and to extract physical information once the waves are observed. EOB models are computationally 
expensive to generate (although far faster than doing NR simulations) and work 
is underway to speed them up. Future work will continue to improve the 
EOB radiation-reaction sector and the calibration of the EOB conservative 
dynamics, extend the modeling to higher-order modes, 
investigate the performance of the model 
against the precessing configurations in the SXS catalog, and check its stability against much longer 
NR simulations, thus extending  the studies recently carried out 
in Ref.~\cite{Pan2013b} for nonspinning BHs. 

{\it Acknowledgments.} A.B., T.H., Y.P. and A.T. acknowledge partial
support from NSF Grants No. PHY-0903631 and No. PHY-1208881. A.B. also
acknowledges partial support from NASA Grant NNX09AI81G. T.H. and
A.T. also acknowledge support from the Maryland Center for Fundamental
Physics.  A.M. and H.P. gratefully acknowledge support from NSERC of
Canada, the Canada Chairs Program, and the Canadian Institute for
Advanced Research.  M.B., D.H., and L.K. gratefully acknowledge
support from the Sherman Fairchild Foundation, and from NSF grants
PHY-1306125 and PHY-1005426 at Cornell. 
M.S., B.S., N.T., and A.Z. acknowledge support from the 
Sherman Fairchild
Foundation and from NSF grants PHY-106881, PHY-1005655, and
DMS-1065438 at Caltech.
G.L. acknowledges 
partial support from NSF Grant No. PHY-1307489.
Simulations used in this work were
computed with the \texttt{SpEC} code~\cite{SpECwebsite}.  Computations
were performed on the Zwicky cluster at Caltech, which is supported by
the Sherman Fairchild Foundation and by NSF award PHY-0960291; on the
NSF XSEDE network under grant TG-PHY990007N; 
on the Orca cluster supported by Cal State Fullerton; 
and on the GPC
supercomputer at the SciNet HPC Consortium~\cite{scinet}. SciNet is
funded by: the Canada Foundation for Innovation under the auspices of
Compute Canada; the Government of Ontario; Ontario Research
Fund--Research Excellence; and the University of Toronto.

\bibliography{References/References}

\end{document}